\documentclass{appolb}
\usepackage{graphicx}
\begin{document}
\title{$\rho$-Meson Form Factors in the Point Form%
\thanks{Presented at the 9th Excited QCD conference, 7-13 May 2017, Sintra, Portugal}%
}
\author{Elmar P. Biernat
\address{Centro de F\' isica Te\' orica de Part\' iculas, Instituto Superior T\'ecnico, Av.\ Rovisco Pais,
1049-001 Lisboa, Portugal}
\\
{Wolfgang Schweiger
}
\address{ Institut f\"ur Physik, Fachbereich Theoretische Physik, Universit\"at Graz, A-8010 Graz, Austria}
}
\maketitle
\begin{abstract}
We present a calculation of the electromagnetic form factors of the $\rho^+$ meson. Our formalism is
based on the point-form of relativistic quantum mechanics. Electron-$\rho$-meson scattering is formulated as a coupled-channel problem for a Bakamjian-Thomas mass operator, such that the dynamics of the exchanged photon is taken explicitly into account. The $\rho$-meson current is extracted from on-shell matrix elements of the optical potential of the scattering process. As a consequence of the violation of cluster separability in the Bakamjian-Thomas framework, our current includes additional, unphysical contributions, which can be separated from the physical ones uniquely. Our results for the form factors are in good agreement with other approaches.
\end{abstract}

\section{Introduction}
For the structure of the simplest hadrons made of two light quarks, there exists a vast amount of experimental data on the pion electromagnetic form factor, but for the $\rho$-meson there is only little data, and only in the timelike region. The experimental difficulty of acquiring data on the $\rho$ meson is due to its very short lifetime. This makes the calculation of the $\rho$-meson form factors of particular theoretical interest. The knowledge of the electroweak properties of the simplest hadrons is of great importance for our understanding of the strong interaction at low energies.

We use point-form relativistic quantum mechanics to calculate the $\rho$-meson electromagnetic current (for a mini-review on the applications of the point form, see Ref.~\cite{Bi10}). The characteristic feature of the point form is the property that only the generators of space-time translations are interaction dependent, whereas the Lorentz generators are free of interactions, with the advantage of keeping Lorentz covariance manifest.\footnote{This property is explicit when one calculates the point-form Poincar\'{e} generators from a given Lagrangian density by integration over a hyperboloid in Minkowski space~\cite{Biernat:2007sz}.}

We employ the so-called \lq\lq Bakamjian-Thomas (BT) construction\rq\rq~\cite{BT53} to introduce interaction terms in the total four-momentum operator by means of a mass operator containing all the dynamics of the problem. One advantage of the BT approach is that it allows
even for instantaneous interactions while maintaining Poincar\'{e} invariance. However, in general, cluster separability (macroscopic
locality) is violated in BT~\cite{KP91}. In this paper we discuss the consequences of this violation on the electromagnetic current and present a solution how to deal with them.

\section{Formalism}
The electromagnetic form factors are the observables that describe the electromagnetic structure of a bound state. Unlike most other theoretical calculations we extract the form factors from the whole physical process in which they are measured, i.e. elastic electron-meson scattering: We formulate this process as a coupled two-channel problem for a BT mass operator~\cite{Bi09}. The two channels, the electron-quark-antiquark ($eq\bar q$) and the electron-quark-antiquark-photon ($eq\bar q\gamma^\ast$) are coupled through vertex operators which are derived from QED~\cite{Kl03}. The natural multi-particle basis states in the point-form BT framework are velocity states characterized by the overall velocity $\vec v$, the center-of-mass momenta $\{\vec k_i\}$, and the canonical spin projections $\{\mu_i\}$ of each particle~\cite{Kl98}.
%\footnote{A peculiar property of the point-form BT framework is the conservation of the overall four-velocity at the interaction vertices.} 
The confining interaction between the quarks is modeled by an (instantaneous) harmonic-oscillator potential which is added to the free invariant $eq\bar q$ and $eq\bar q\gamma^\ast$ mass operators. From matrix elements of the one-$\gamma^\ast$-exchange optical
potential between velocity states, obtained from the coupled-channel problem by a Feshbach reduction, we can extract the matrix elements of the electromagnetic $\rho$-meson current as~\cite{Bi14}
\begin{eqnarray}
&&J^\mu(\vec{k}^\prime_\rho,\mu_\rho^\prime ; \vec{k}_\rho,\mu_\rho)=
\sqrt{\omega_{k_\rho}\omega_{k_\rho^{\prime}}}\sum_{\mu_q'\mu_{\bar q}'
} \int \frac{\mathrm d^3 k_{q}'}{\omega_{k_q'}}\sqrt{\frac{\omega_{\tilde k_q'} \omega_{\tilde k_{\bar q}'}}{\omega_{\tilde k_q'}+\omega_{\tilde k_{\bar q}'}}}\sqrt{\frac{\omega_{\tilde k_q} \omega_{\tilde k_{\bar q}}}{\omega_{\tilde k_q}+\omega_{\tilde k_{\bar q}}}} \nonumber\\&&\times\frac{\sqrt{ \omega_{k_{q}'}+\omega_{k_{\bar q}'}}}{\omega_{k_{\bar q}'}}\frac{\sqrt{\omega_{k_{q}}+\omega_{k_{\bar q}}}}{\omega_{k_q}} \sum_{\mu_q} \psi^\ast_{\mu'_\rho\mu_q'\mu_{\bar{q}}'}(\vec{\tilde k}_q') \, \psi_{\mu_\rho\mu_q\mu_{\bar{q}}'}(\vec{\tilde k}_q)
 \, Q_q\,  j_q^\mu(\vec{k}_q^\prime,\mu_q^\prime;\vec{k}_q,\mu_q)\nonumber\\&&+ (q\leftrightarrow \bar q) \,\, ,  \label{eq:Jalpha}
 \end{eqnarray}
where $\omega_{k_i}=\sqrt{m_i^2+\vec k_i^2}$, $\vec k_i^{(\prime)}$ is the incoming (outgoing) three-momentum of particle $i$ in the overall rest frame, a tilde on the top denotes a momentum in the meson rest frame, $\psi$ is the meson wave function, $j_q^\mu$ is the quark current, and $Q_q$ is the quark charge in units of the elementary charge $\mathrm e$.

It turns out that there are 11 independent spin matrix elements of the current~(\ref{eq:Jalpha}), which means that one needs 11 form factors to fully parametrize it. These are the usual three physical form factors $f_1$, $f_2$, and $g_M$, and additional 8 spurious (unphysical) form factors, associated with covariants proportional to the sum and difference of incoming and outgoing electron four-momenta.\footnote{Notice that some of these structures violate current conservation, however covariance and hermiticity of the current are still satisfied.} Furthermore, all form factors depend, in addition to the momentum transfer squared $Q^2$ (Mandelstam $t$), also on Mandelstam $s$, the total invariant mass squared of the electron-meson system. The reason for the appearance of these spurious dependencies is the violation of cluster separability in the BT framework. A numerical analysis shows that some spurious contributions vanish for large $s$, which suggests to take the limit $s\rightarrow\infty$. The resulting current resembles the one obtained within a covariant
light-front approach~\cite{Carbonell:1998rj} and the physical form factors can be extracted in an unambiguous and clean way, without any spurious contributions. In particular, for a momentum transfer in the $x$-direction we have~\cite{Bi14}
\begin{eqnarray}
&&F_1(Q^2)= \lim_{s\rightarrow\infty} f_1(Q^2,s)=
\lim_{s\rightarrow\infty}\frac{(-1)}{\sqrt s}\left[J^0(\vec{k}_\rho^\prime,1;
\vec{k}_\rho,1)+J^0(\vec{k}_\rho^\prime,1;
\vec{k}_\rho,-1)\right], \nonumber\\
&&F_2(Q^2)=\lim_{s\rightarrow\infty}
f_2(Q^2,s)=\lim_{s\rightarrow\infty}\frac{(-4m_\rho^2)}{Q^2\sqrt s}J^0(\vec{k}_\rho^\prime,1;
\vec{k}_\rho,-1),\nonumber\\
&&G_M(Q^2)=\lim_{s\rightarrow\infty}
g_M(Q^2,s)=\lim_{s\rightarrow\infty}\frac{(-\mathrm
i)}{Q}J^2(\vec{k}_\rho^\prime,1; \vec{k}_\rho,1),
\end{eqnarray}
where
\begin{eqnarray}
 &&\lim_{s\rightarrow\infty}\frac{J^0(\vec{k}_\rho^\prime,1;
\vec{k}_\rho,\pm1)}{\sqrt s} =\frac{1}{4\pi}\int\mathrm{d}^3\tilde{k}'_q\sqrt{\frac{m_{q\bar q}}{m'_{q\bar q}}}u^\ast\left(|\vec{\tilde{k}}_q'|\right)u\left(|\vec{\tilde{k}}_q|\right)\mathcal S_{1\pm1}^+\,,\nonumber\\
&&\lim_{s\rightarrow\infty}{J^2(\vec{k}_\rho^\prime,1;
\vec{k}_\rho,1)} \nonumber\\&&=\frac{1}{4\pi }\int\mathrm{d}^3\tilde{k}'_q\sqrt{\frac{m_{q\bar q}}{m'_{q\bar q}}}u^\ast\left(|\vec{\tilde{k}}_q'|\right)u\left(|\vec{\tilde{k}}_q|\right)
\frac{m_{q\bar q}'}{ ( m_{q\bar q}' + 2  \tilde k_q'^3)}
\left(
\tilde k_q'^2 \mathcal S_{11}^{+}+  \frac{\mathrm i \,Q}{2}
\mathcal S_{11}^{-}\right),\nonumber\\
\end{eqnarray}
with $m_{q\bar q}$ the free invariant mass of the $q\bar q$ system, $u$ the radial $\rho$-meson wave function, and $\mathcal S^{\pm}_{\mu_\rho'\mu_\rho}$ the spin rotation factors given by
\begin{eqnarray}\label{eq:S1}
\mathcal S^{\pm}_{\mu_\rho'\mu_\rho}&&=\lim_{s\rightarrow \infty}\frac12\sum_{ \mu_q, \tilde\mu_q,\ldots}(\pm 1)^{\mu_q-\frac12} D^{\frac12}_{\mu_q\tilde\mu_q}\left[ R_W\left(\frac{\tilde k_q}{m_q},B_c\left(v_{q\bar q} \right)\right)\right]\left(\vec{\epsilon}_{\mu_\rho}\cdot \vec{\sigma}\right)_{\tilde\mu_q \tilde\mu_{\bar q}}
\nonumber\\&&\times D^{\frac12}_{\tilde\mu_{\bar q}\tilde\mu_{\bar q}'}\left[R_W\left(\frac{\tilde k_{\bar q}'}{m_q},B_c^{-1}\left(v_{q\bar q}\right)B_c\left(v'_{q\bar q}\right)\right)\right]  \left(\vec\epsilon^{\ast}_{\mu_\rho'}\cdot \vec \sigma\right)_{\tilde\mu_{\bar q}' \tilde\mu_{q}'}\nonumber\\&&\times D^{\frac12}_{\tilde\mu_q'\mu_q}\left[R_W\left(\frac{k_q'}{m_q},B_c^{-1}\left(v'_{q\bar q} \right)\right)\right]\,\,.
\end{eqnarray}
Here the $D$'s are the usual Wigner $D$-functions depending on a  Wigner rotation $R_W$ associated with a canonical (rotationless) boost $B_c$,  $v_{q\bar q}=(k_q+k_{\bar q})/m_{q\bar q}$ is the four-velocity of the free $q\bar q$ system, and $\vec\epsilon_{\mu_\rho}$ are the spin-1 meson polarization three-vectors in the rest frame.

\section{Results}

\begin{figure}[htb]
\centerline{%
\includegraphics[width=8.6cm]{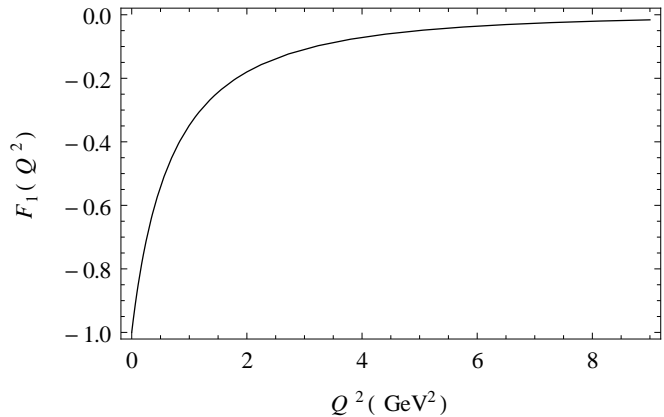}}
\caption{$F_1\left(Q^2\right)$ with parameters $m_q=0.34\, \mathrm {GeV}$, $a=0.312\, \mathrm {GeV}$ and $m_\rho= 0.77\, \mathrm {GeV}$.}
\label{fig:rhoFF1}
\end{figure}

\begin{figure}[htb]
\centerline{%
\includegraphics[width=8.3cm]{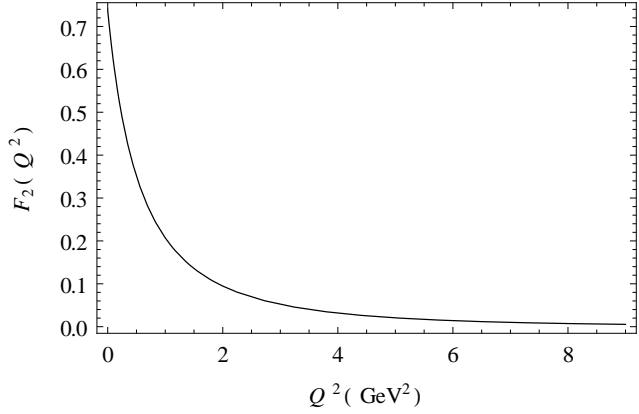}}
\caption{$F_2\left(Q^2\right)$ with the same parameter values as in Fig.~\ref{fig:rhoFF1}.}
\label{fig:rhoFF2}
\end{figure}

\begin{figure}[htb]
\centerline{%
\includegraphics[width=8.3cm]{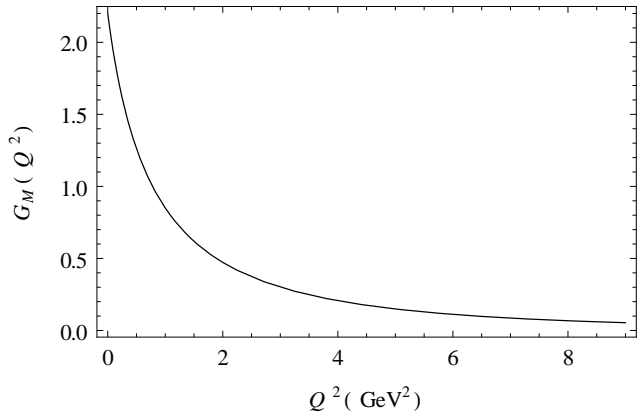}}
\caption{  $G_M\left(Q^2\right)$ with the same parameter values as in Fig.~\ref{fig:rhoFF1}.}
\label{fig:rhoGM}
\end{figure}

For the numerical computation of the form factors we take a simple harmonic-oscillator wave function for the $u$. It depends on the oscillator length $a$, which is fixed, together with the quark mass $m_q$, by the ground state (the $\rho$-meson with $m_\rho=0.77$ GeV) and the first two radial excitations of the vector-meson spectrum.
For the values fixed at $m_q= 0.34$ GeV and $a =
0.312$ GeV the form factors are depicted in Figs. \ref{fig:rhoFF1}-\ref{fig:rhoGM}.

Our predictions for the magnetic dipole and the electric quadrupole moment, are $\mu_\rho=2.2$ and $Q_\rho=-0.47$ (in units $|\,e\,|/2m_{\rho}$ and $|\,e\,|/m_{\rho}^2$), respectively. These predicted values lie between those of most other approaches. Particularly with the covariant light-front approach of~\cite{Carbonell:1998rj} we find excellent agreement when the same wave functions and parameters are used in both calculations.

\section{Conclusions and outlook}

In the present work we have applied our previously developed point-form coupled-channel formalism to the calculation of the $\rho$-meson electromagnetic form factors. As in the spin-0 pion case we also find for the spin-1 $\rho$-meson remarkable similarities with the covariant light-front approach of Carbonell et al.~\cite{Carbonell:1998rj}.
Current and future applications of our quite general formalism include deuteron form factors~\cite{Biernat:2010py}, electroweak form factors of heavy-light mesons~\cite{GomezRocha:2011qs,GomezRocha:2012zd,Senekowitsch:2013vha}, nucleon form factors and pion-cloud effects~\cite{Kupelwieser:2015ioa}, baryon transition form factors such as N-$\Delta$ transition form factors~\cite{Ju:inprep}, and meson-baryon vertices~\cite{Jung:2017cpy}.

\subsection*{Acknowledgements}
This work received financial support from the \lq\lq Fonds zur F\"{o}rderung der wissenschaftlichen Forschung in \"{O}sterreich'' under grant No. FWF DK W1203-N16, as well as from the Province of Styria, Austria under a PhD grant. This work was also partially supported by the \lq\lq Funda\c c\~ao para a Ci\^encia e a Tecnologia (FCT)'' under grant Nos.~PTDC/FIS/113940/2009, CFTP-FCT (UID/FIS/0777/2013), SFRH/BPD/\-100578/\-2014, and by the European Union under the HadronPhysics3 Grant No. 283286.

%
%
% \section{Next section}
% The text...
% \subsection{Subsection}
% The text...

%uncomment the following lines to place a figure
%\begin{figure}[htb]
%\centerline{%
%\includegraphics[width=12.5cm]{Fig1}}
%\caption{Plot of ...}
%\label{Fig:F2H}
%\end{figure}

\end{document}